\documentclass[11pt]{revtex4}

\usepackage{graphicx}
\usepackage{newtxtext}
\usepackage{newtxmath}
\usepackage{natbib}
\usepackage{hyperref}
\hypersetup{
    colorlinks = true,
    urlcolor   = blue,
    citecolor  = black,
}

\newcommand{\RomanNumeralCaps}[1]
\linenumbers

\begin{document}

\title{Relative accelerations characterize the hydrodynamic interaction of cloud droplets}

\author{Reece V. Kearney} 
\author{Gregory P. Bewley$^*$} 

\affiliation{
Sibley School of Mechanical and Aerospace Engineering, Cornell University, Ithaca, NY, 14853, USA \\
$^*$email for correspondance: gpb1@cornell.edu}


\begin{abstract}
Water droplets coalesce into larger ones in atmospheric clouds to form rain. 
But droplets on collision courses do not always coalesce due to the cushioning effects of the air between them. 
The extent to which these so-called hydrodynamic interactions reduce coalescence rates is embodied in the collision efficiency, 
which is often small and is not generally known. 
In order to characterize the mechanisms that determine the collision efficiency, 
we exploited new time-resolved three-dimensional droplet tracking techniques to measure the positions of cloud droplet pairs settling through quiescent air. 
We did so with an unprecedented precision that enabled us to calculate 
the relative positions, velocities, 
and accelerations of the droplets 
at droplet surface-to-surface separations as small as about one-tenth of a droplet diameter. 
We show that relative accelerations clearly distinguish coalescing from non-coalescing droplet trajectories, 
the former being associated with relative accelerations that exceeded a threshold value. 
We outline how relative accelerations relate to hydrodynamic interactions, 
and present scaling arguments that predict the threshold relative acceleration. 
We speculate that the relative acceleration distribution of droplets in turbulent clouds can parameterize the collision efficiency, 
and that this distribution together with the well-known relative position and velocity distributions 
can generate a physical description of both the collision and coalescence rates of cloud droplets. 
\end{abstract}


\maketitle


\section{Introduction}
\label{sec:intro}

In warm atmospheric clouds, 
liquid water droplets are set onto collision courses by mechanisms that include differential sedimentation and turbulence 
\cite[][]{shaw2003,devenish2012,grabowski2013}. 
The rate at which this happens can be explained by the way turbulence generates droplet relative position and velocity distributions 
\cite[][]{sundaram1997,chun2005}. 
In order for droplets on collision courses to coalescence into one larger droplet, however, 
they need to squeeze out the air between them, 
a process that couples the motions of droplets relative to one another through hydrodynamic interaction (HI) 
\cite[\textit{e.g.}][]{pinsky2007,how2021}. 
These interactions cause droplets to decelerate, 
which modifies the probability distributions of droplet relative positions and velocities 
in a way as yet unknown 
\cite[\textit{e.g.}][]{yavuz2018,hammond2021,bragg2022}. 
The effect is most pronounced precisely where these distributions need to be evaluated, 
which is at the moment of contact between two droplets. 

	Theoretical and empirical descriptions of the relative velocity distributions \cite[\textit{e.g.}][]{saw2014,gustavsson2014,meibohm2017}
and radial distribution functions \cite[\textit{e.g.}][]{reade2000,bec2007,larsen2018} 
exist for droplets that are separated widely enough that they do not interact with one another. 
One approach to incorporate HI is to modify our descriptions of the relative velocities and radial distributions to account for HI 
\cite[\textit{e.g.}][]{yavuz2018,bragg2022}. 
For the purposes of this paper, we instead consider drawing values for the relative velocities and radial distributions 
from the relatively mature theories for noninteracting inertial particles in turbulence, 
and to modify the corresponding collision-rate prediction with a multiplicative collision efficiency 
that incorporates a physical understanding of the mechanisms by which HI reduces coalescence rates
\cite[\textit{e.g.}][]{klett1973,wang2005}. 

	The outcome of binary droplet interactions can be parameterized by the impact parameter, 
which quantifies the offset from a head-on collision of the droplets' trajectories \cite[\textit{e.g.}][]{ashgriz1987,qian1997}. 
A challenge this parameterization faces in turbulent settings 
is that the impact parameter is defined asymptotically far ahead in time of a potential collision. 
In any practical setting, the conditions change in the far-field, 
making the impact parameter difficult to measure in an experiment, 
even in one as well controlled as the one under scrutiny in this paper. 
Since droplets do move against a backdrop of turbulent flow in clouds, 
and the impact parameter is hard to identify, 
we propose and explore an alternative parameterization in terms of the relative accelerations of the droplets, 
which to our knowledge is new. 

	We quantified the near-contact motions of freely-falling cloud droplets 
with a precision that was for the first time sufficient to measure the relative accelerations of the droplets 
caused by their hydrodynamic interactions. 
Previous experimental studies have reliably quantified individual droplet positions, velocities, and accelerations 
\cite[\textit{e.g.}][]{voth2002,ayyalasomayajula2006,good2014}, 
or the relative positions and velocities (only) of droplet pairs separated by tens of diameters or larger 
\cite[\textit{e.g.}][]{xu2011,bewley2013,yavuz2018,hammond2021}. 
Here we evaluated these quantities and furthermore calculate the relative acceleration 
down to surface-to-surface separations as small as one-tenth the droplets' diameters as described below. 
In Sec.\,\ref{sec:theory} we outline scaling arguments that capture the essential features 
of the way relative accelerations relate to HI and coalescence. 
We describe in Sec.\,\ref{sec:methods} how we analyzed the data, 
and in Sec.\,\ref{sec:results} how the data are consistent with our scaling arguments. 
We draw particular attention to the observation that coalescing droplet pairs experienced larger accelerations 
than non-coalescing pairs, 
and discuss how the relative acceleration distribution in turbulent environments 
might complement the relative position and velocity distributions to model droplet coalescence rates.

\begin{figure}
  \centerline{\includegraphics[width=0.25\textwidth]{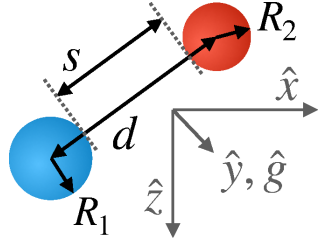}}
  \caption{Pairs of droplets with radii $R_1 \gtrsim R_2 \approx 20\,\mu m$ 
  fell in the $\hat{y}$-direction, the direction of gravity, 
  and we examined the evolution of their center-to-center and surface-to-surface separations, 
  $d$ and $s$ respectively, 
  and the time derivatives thereof at separations as small as $0.1R$. 
  In this figure and in the next, gravity points predominantly out of the page. }
\label{fig:coords}
\end{figure}

\section{Theory}
\label{sec:theory}
Consider two droplets that approach each other with an initial relative velocity, $\delta v_0$, 
and impact parameter, $B \equiv \chi /R$, 
where the collision radius, $R = R_1 + R_2$, is the sum of the radii of the two droplets, 
and $\chi$ is the component of the center-to-center separation vector that is orthogonal to the relative velocity of the droplets 
when they are asymptotically far apart. 
When $B=0$ a head-on collision would result, and for $B=1$ a grazing collision, if the droplets did not interact. 
The intervening air causes a hydrodynamic interaction (HI) that prevent collisions above a certain critical value, $0 \leq B_c < 1$. 
The particular value of the critical impact parameter, $B_c$, depends on the characteristics of the droplets and the carrier fluid (air), 
and is approximately 0.3 for cloud droplets with radii near 20\,$\mu$m \cite[][]{pruppacher2012}. 
To simplify the problem, consider that the air surrounding the droplets is quiescent apart from the disturbances caused by the droplets' motions. 

	Hydrodynamic interaction dissipates the relative velocities of droplets. 
For coalescence to occur, the radial relative velocity at contact, $\delta v_c$, 
must be greater than or equal to zero despite the deceleration, such that $\delta v_c \geq 0$. 
We identify a marginal deceleration that divides initial conditions that lead to coalescence from those that do not 
as the relative acceleration, $\delta a_c$, associated with $\delta v_c = 0$. 
In the marginal case where the final relative velocity is zero, 
we identify a characteristic acceleration given by $\delta a_c = \delta v_0^2/S$, 
where $S$ is the distance over which the HI operates. 
For droplets settling freely under gravity, 
the initial relative velocity is the difference between their terminal settling speeds 
so that $\delta v_0 = \delta v_g$. 
We revisit this condition below, after noting the existence of an internal velocity scale as follows. 

	The interaction between two droplets can be characterized by an internal velocity scale, 
which we see by considering small separation distances, 
$s = d-R$, 
where $d$ is the center-to-center distance between the droplets. 
Ignoring rotations in this regime, lubrication theory predicts that the forces between the droplets 
are proportional to $\mu R^2 \delta v / s$, 
noting a characteristic $1/s$ dependence \cite[\textit{e.g.}][]{batchelor2000}. 
Given that $\delta a dt = d \delta v$ and $\delta v dt = ds$, 
momentum conservation takes the form $d \delta v = -C_\nu (p/M) ds/s$, 
where $p = \mu R_1 R_2$ is a viscous momentum scale, 
$M = m_1m_2 / (m_1 + m_2)$ is the harmonic mean mass of the two droplets, 
and $C_\nu = 6 \pi$ according to lubrication theory. 
Under these circumstances the relative velocity decays logarithmically with distance \cite[][]{li2021} so that 
\begin{equation}
\delta v = -\delta v_i \log(s/s_0), 
\label{eq:dv}
\end{equation}
and the viscous deceleration of droplet pairs operates with a characteristic velocity, 
$\delta v_i = C_\nu p/M$, 
which depends only on the characteristics of the droplets and fluid, and not on their initial velocities. 
We redefine the marginal relative acceleration with this internal velocity to be $\delta a_c \equiv \delta v_i^2 / R$ 
under the assumptions that 
\textit{(i)} the interaction is governed by $\delta v_i$ and not by the  
initial relative velocity, 
and \textit{(ii)} that the interaction distance scales with the droplet sizes such that $S = R$. 
Finally, we observe that the relative acceleration of a given droplet pair, $\delta a_c$, 
equals the marginal one when 
\begin{equation}
{A_c}  \equiv  \frac{\vert \delta a \vert M^2}{\mu^2 R^3} 
= C_\nu^2 / 16
\label{eq:dac}
\end{equation}
at leading order for droplets of nearly the same size, 
and where $A_c$ takes a value of about one according to lubrication theory. 
We expect relative accelerations to exceed the critical one only when droplets coalesce. 
When hydrodynamic interactions are so strong that they cushion the droplets and prevent coalescence, 
the droplets then roll off one another, and the relative accelerations decrease 
and then change sign as the droplets separate. 

Equation \ref{eq:dac} predicts a marginal relative acceleration 
that is exceeded only by droplets that coalesce, 
which is a threshold that is determined by intrinsic parameters ($\mu$, $R$, $M$ and $C_l$) 
and not by the droplets' initial relative velocities. 
If the initial relative velocity were to dominate the interaction, 
then the marginal relative acceleration would be $\delta a_c = \delta v_g^2/R$ 
for droplets initially settling at their terminal velocities. 
The normalized relative acceleration, $A_c$, then takes a value proportional to $g^2 M^4 / \mu^4 R^6$ 
and to the size \textit{difference} between droplets in a pair. 
In the experiments described below, 
this value of $A_c$ is $O(10^3)$, 
to be compared with $O(1)$ for $A_c$ determined by the interaction velocity, $\delta v_i$ as described above. 
In Sec.\,\ref{sec:results} we examine the extent to which the approximations outlined above held in the experiments, 
including whether the inter-droplet forces decayed as $1/s$, whether the logarithmic decay of velocity prevailed, 
and whether the relative acceleration was governed by intrinsic or extrinsic factors in the experiments.


%

\begin{figure}
  \centerline{\includegraphics[width=0.55\textwidth]{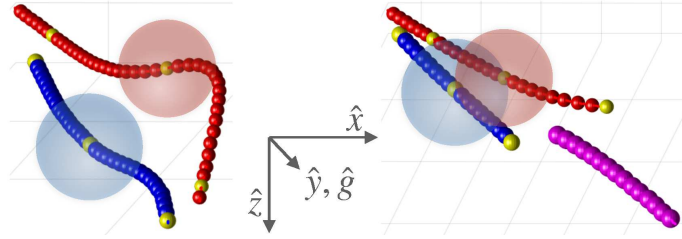}}
  \caption{The larger blue droplet fell down (predominantly out of the page) 
  onto the smaller red one 
  and pushed it out of the way while squeezing out the air between the two. 
  The cushioning effect of HI, 
  causes the trajectories to curve and to 
  appear to separate from each other in space, though the distance between the droplets decreased monotonically in time. 
  If there were no HI the trajectories would have been straight. 
  The droplets settled 
  by about 2\,mm 
  with every seventh position marked by a sphere that is scaled down by a factor of eight for clarity (the translucent red and blue spheres are to scale). 
  The yellow spheres are colocated in time. 
  \textit{Left:} The blue droplet fell past the smaller red one without touching it. 
  \textit{Right:} Under similar conditions but with a slightly smaller initial impact parameter, 
  the blue and red droplets coalesced to form the pink one. }
\label{fig:trajectories}
\end{figure}

\section{Methods}
\label{sec:methods}
	We analyzed data described in \cite{ivanov2017} and \cite{magnusson2022}, 
and studied interactions between 186 pairs of water droplets settling past each other in still air as illustrated in Fig.\,\ref{fig:coords}. 
The data are the subset of those analyzed in \cite{magnusson2022} for which the charges were approximately zero 
and the initial relative velocities were close to their terminal value for non-interacting droplets. 
Subsequent deviations from constant relative velocities were due to HI, 
which can be seen in part in the curvature of the trajectories through space in Fig.\,\ref{fig:trajectories}. 
The droplet radii were between 17 and 27\,$\mu$m, 
putting them in the size gap of cloud droplet growth \cite[\textit{e.g.}][]{shaw2003}. 
The size ratio, $\kappa = R_2 / R_1$, between interacting droplets varied within about 20\% of one. 
The initial droplet separations were typically about ten droplet diameters. 
The initial impact parameters were defined by the separation and relative velocity when droplets were five collision radii apart. 
The distribution of impact parameters, $0 < B \lesssim 2$, 
was determined by the peculiarities of the experiment and differs from the one arising in turbulent clouds. 
While fluctuations in the atmosphere induce interactions between droplets within evolving background flows, 
the special case of approach induced by differential sedimentation in otherwise quiescent air also occurs in clouds 
and constitutes a benchmark for the understanding of coalescence rates in more complex settings. 
	
\subsection{Methods to track droplet motions}
Two high-speed cameras with long-range microscopes recorded droplet shadows 
with a resolution of approximately 2\,$\mu$m per pixel at a frame rate of 8.5\,kHz \cite[][]{magnusson2022}. 
The observation volume was about 0.5\,mm $\times$ 5\,mm $\times$ 2\,mm, with the long axis aligned with gravity. 

	Our custom Lagrangian droplet tracking algorithm generated three-dimensional trajectories for individual droplets 
and did so even when the droplets were separated by only a fraction of their diameter and approaching contact \cite[][]{kearney2020}. 
The algorithm automatically detected low Weber number binary collisions that resulted in coalescence, 
and in doing so improved on previous methods by not requiring human intervention \cite[\textit{e.g.}][]{aarts2005,bordas2013}. 
Collisions were identified by a combination of criteria, including conservation of droplet mass and momentum \cite[][]{kearney2020}. 
Existing automated tracking methods worked only for droplets that were far apart and did not merge or split apart \cite[\textit{e.g.}][]{qian1997}. 
Our methods identified particles that overlapped in projection by over 80\% even in noisy images. 
We measured the sizes and positions of droplets on each camera using Pratt-Walking, 
which found the sub-pixel center locations of overlapping circles in digital images by using edge data \cite[][]{kearney2020}. 
Droplet radii were calibrated by particles of known size in order to account for systematic errors stemming from blurring and thresholding. 
We then tracked the droplets in two-dimensions separately on each camera using a hybrid method \cite[][]{ouellette2006,kearney2020} 
and reconstructed three-dimensional trajectories by matching droplets between the cameras based on their sizes. 

\subsection{Methods to characterize droplet motions}
\label{subsec:motionmethods}
From droplet trajectories in three dimensions, 
we calculated velocities and accelerations by filtered finite differences \cite[][]{kearney2020}. 
Noise was suppressed in the accelerations with polynomial fits to the data. 
We estimated the uncertainty in acceleration measurements with synthetic data. 
From droplet trajectories given by the model in \cite{jeffrey1984}, 
we generated synthetic images of droplets with a quartic intensity profile and with additive Gaussian noise. 
Pairs of images that mimicked the views and resolutions of the cameras in the experiments were generated. 
We found that the error in measuring the relative radial acceleration was less than 2.0\,$\times$\,$10^{-3}$\,pixels/frame$^2$ 
in the sense of a 95\% confidence interval, 
which corresponds to an uncertainty of 0.26\,m/s$^2$ at the spatial resolution and sampling frequency of the experiment, 
or to an uncertainty of about 3\% of the gravitational acceleration. 

\subsection{How relative accelerations characterize droplet interaction forces}
The relative accelerations of droplet pairs are related to hydrodynamic interaction forces as follows. 
We decompose the sum of the forces on each droplet (labeled $\gamma$ = 1 and 2) 
into the one component imposed by the surrounding fluid, $\mathbf{F}_{f,\gamma}$, 
and one by gravity so that $\mathbf{F}_\gamma = \mathbf{F}_{f,\gamma} + m_\gamma \mathbf{g}$, 
where $\mathbf{g}$ is the acceleration of gravity and the bold font indicates a vector quantity. 
According to Newton's Second Law the forces and accelerations are proportional, 
$\mathbf{F}_\gamma = m_\gamma \mathbf{a}_\gamma$, 
so that the relative accelerations, $\delta \mathbf{a}$, 
are given by 
\begin{equation}
\delta \mathbf{a} \equiv \mathbf{a}_1 - \mathbf{a}_2 
= \mathbf{F}_{f,1}/m_1 - \mathbf{F}_{f,2}/m_2, 
\label{eq:dam}
\end{equation}
where the contributions from gravity cancel out. 
When the forces on the droplets are equal and opposite, $\delta \mathbf{a} = \mathbf{F}_{f}/M$, where $M$ is defined in Sec.\,\ref{sec:theory}. 
This special case is approached as the droplet separation vanishes. 
We confine our interest to the radial relative accelerations, $\delta a_{r} \equiv \delta \mathbf{a} \cdot \mathbf{\hat{r}}$, 
where $\mathbf{\hat{r}}$ is the unit vector aligned with a vector connecting the droplet centers, 
since it is the radial component of the relative acceleration that characterizes the change in the rate at which droplets approach collision. 
The fluid forces on each droplet, $\mathbf{F}_{f,\gamma}$, are known in different limiting circumstances, 
which we consider separately as needed.

\begin{figure}
  \centerline{\includegraphics[width=0.7\textwidth]{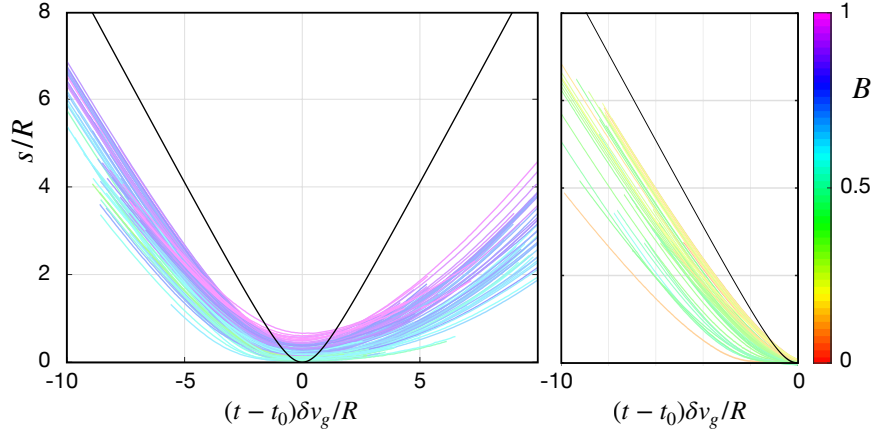}}
  \caption{The surface-to-surface distances ($s$) between non-coalescing (\textit{Left}, $N$ = 123) 
  and coalescing (\textit{Right}, $N$ = 63, in a separate graph for clarity) 
  droplet pairs decreased to a minimum at $t_0$. 
  The impact parameter ($B$, line color) was evaluated when droplets were five collision radii ($R$) apart. 
  In the absence of HI, coalescence would occur for $B < 1$. 
  In these experiments, coalescence occurred when $B < B_c \approx 0.5$. 
  A synthetic noninteracting droplet pair (black line) for which $\delta v = \delta v_g$ is constant 
  illustrates the grazing trajectory ($B$ = 1). }
\label{fig:separations}
\end{figure}

\section{Results and discussion}
\label{sec:results}
	Two representative events, a near-miss and a coalescence, are in Fig.\,\ref{fig:trajectories} 
with the smaller droplet (red) initially leading the larger droplet (blue). 
In these views the droplets settled gravitationally out of the page primarily, 
with a small component from top-left to bottom-right of the image. 
In the left panel, the droplets displaced each other as the larger droplet overtook the smaller one before leaving the observation volume. 
For the event on the right, the droplets were unable to move out of each others' way, 
so that they contacted and coalesced into the droplet shown in pink. 
In the absence of HI, the trajectories would have been straight, 
and the curvature of the trajectories is a clear sign of HI, 
which was strongest when the droplets were closest together. 
	
	Figure \ref{fig:separations} shows that droplet pairs spent more time close together than they would have in the absence of HI, 
consistent with an increase in the radial distribution function of droplets in turbulence due to HI 
\cite[\textit{e.g.}][]{yavuz2018,bragg2022}, 
albeit only within distances on the order of the droplet diameter. 
Ballistic trajectories, exemplified by the black line in Fig.\,\ref{fig:separations}, 
correspond to no change in the radial distribution function. 
The observed HI effect corresponds to an increased likelihood of finding droplets within small distances of one another. 
For instance, droplets typically spent more than ten units of time within two diameters of each other 
compared with about five units of time in the same vicinity if they had settled at a constant relative velocity, $\delta v_g$, 
equal to the difference between their terminal settling speeds. 
The effect is accentuated by the fact that droplets departed from each other more slowly than they approached, 
which is evident in the smaller slope of the data for $t > t_0$ than for $t < t_0$. 
The asymmetry in time supports the idea that the radial distribution function relevant to geometric collision rate computations 
is the one conditioned on approaching droplets and excluding droplets moving away from one another. 
This effect is more easily seen in the relative velocities themselves, considered next.

\begin{figure}
  \centerline{\includegraphics[width=0.45\textwidth]{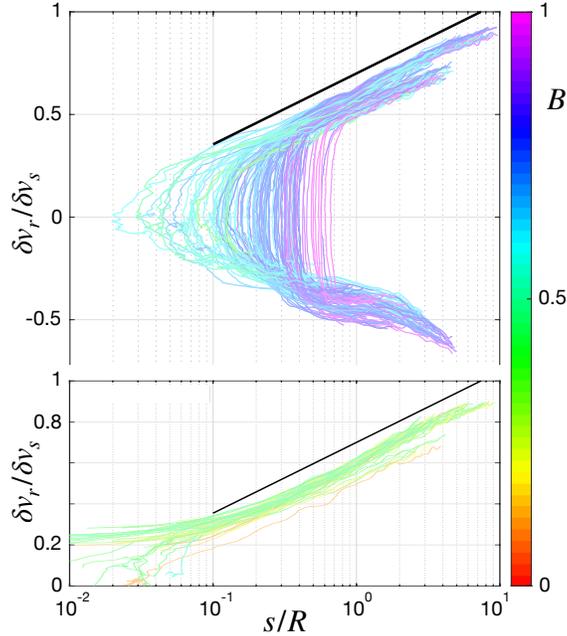}}
  \caption{The radial relative velocities ($\delta v_{r}$) between non-coalescing (\textit{Top}, $B > B_c$) 
  and coalescing (\textit{Bottom}, $B < B_c$) 
  droplet pairs decreased approximately as a logarithm (black lines) of the interfacial distance between them ($s$), 
  consistent with HI (\textit{e.g.} Eq.\,\ref{eq:dv}). 
  The data and color scheme are the same as in Fig.\,\ref{fig:separations}. 
  Dissipation by HI caused droplets to separate from each other more slowly than they approached. 
  Even in the absence of HI, 
  radial relative velocities decrease due to changes in the angle between the relative velocity and the line connecting the droplet centers, 
  a geometric effect captured in $\delta v_s$ as described in the text. }
\label{fig:velocities}
\end{figure}

	The dissipative nature of HI can be seen in the droplet relative velocities shown in Fig.\,\ref{fig:velocities}, 
which were smaller than they would have been in the absence of HI 
consistent with narrower droplet relative velocity distributions in turbulence due to HI 
\cite[\textit{e.g.}][]{yavuz2018,bragg2022}. 
The radial component of the relative velocity, $\delta v_r$, 
is of interest since it determines the rate at which droplets approach a collision. 
To account for the rotation of the separation vector as droplets pass one another by, 
we normalized the relative velocity by the characteristic velocity of freely settling droplets, $\delta v_s$, 
which is given by 
\begin{equation}
\delta v_s \equiv - \delta v_g \sqrt{1-B^2R^2/d^2}
\label{eq:dvs}
\end{equation}
which is equal in magnitude to $\delta v_g$ at large separations, 
and which would be zero if the center-to-center separation, $d$, were equal to $BR$ (which it never was in the experiments). 

	Consistent with Eq.\,\ref{eq:dv}, 
the relative velocity decreased approximately as a logarithm of the surface-to-surface separations 
within an interval of about an order of magnitude centered on $s = R$. 
The droplets' relative velocities were dissipated by HI even at relatively large separations of about $10R$, 
so that the droplets approached one another more slowly than in the absence of HI. 
Since the reduction is logarithmically slow in $s$ the decrease in relative velocity was not pronounced until $s \lesssim 0.1R$. 
Droplets that coalesced typically did so with a relative velocity that saturated toward a constant value at small separations. 
On the other hand, when no coalescence occurred the relative velocity changed sign 
and the droplets departed from one another more slowly than they approached 
since HI is dissipative and opposes changes in both incoming and outgoing relative velocities. 
The changes in slope correspond to a decreased variance and an increased skewness in relative velocities for separation distances 
within about one diameter. 
This points to the importance of finding mean incoming relative velocities to compute geometric collision rates in practice. 
	
\begin{figure}
  \centerline{\includegraphics[width=0.8\textwidth]{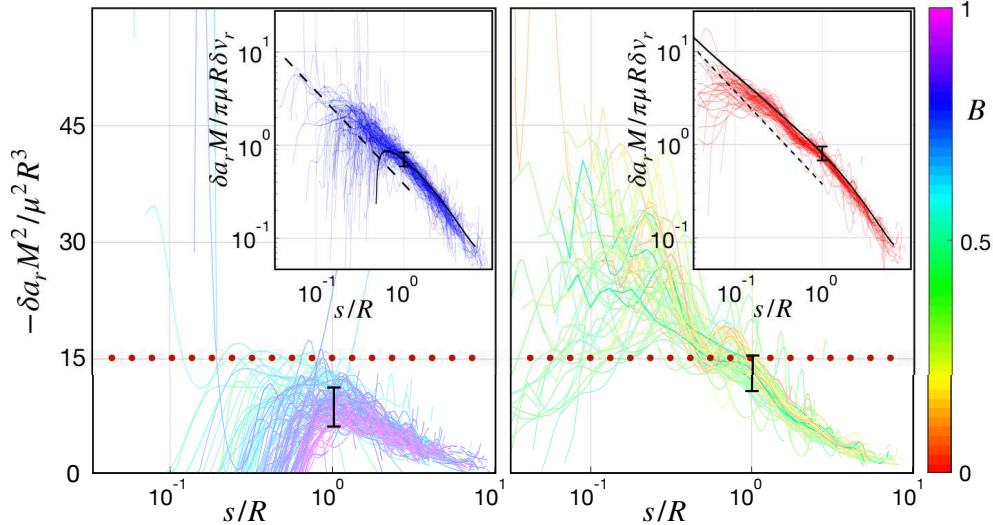}}
  \caption{The radial relative accelerations ($\delta a_{r}$) between non-coalescing (\textit{Left}, $B > B_c$) 
  and coalescing (\textit{Right}, $B < B_c$) 
  droplets increased as the distance between them decreased, 
  and only exceeded a threshold value (of about 15) when the droplets coalesced. 
  The main figures relate to Eq.\,\ref{eq:dac} and the insets relate to lubrication theory (dashed lines, Sec.\,\ref{sec:theory}). 
  The insets also facilitate comparisons with the model in \cite{jeffrey1984} for $B = 1$ (black curve, \textit{Left}) 
  and $B = 0$ (black curve, \textit{Right}). 
  The data and color scheme for the main figures are the same as in Fig.\,\ref{fig:separations}; 
  for the insets all coalescing and non-coalescing trajectories are grouped as red and blue curves, respectively. 
  In the absence of HI, the relative accelerations would be zero. }
\label{fig:accelerations}
\end{figure}

	Figure \ref{fig:accelerations} (main figures) shows that 
only coalescing droplet pairs experienced 
relative accelerations exceeding a threshold value. 
Empirically, we find that a value of about 15 
separates non-coalescing droplet pairs from coalescing ones in the interval 
$0.1 \lesssim s/R \lesssim 1$. 
At yet smaller separations, the data are too sparse to draw conclusions, 
while at larger separations the relative accelerations were so small that the relative uncertainty is too large 
to clearly differentiate between coalescing and non-coalescing pairs. 
The insets compare the relative accelerations with the relative velocities 
in order to expose the inverse power law in $s$ predicted by lubrication theory (Sec.\,\ref{sec:theory}). 
The relative velocities vanish where $s/R$ reaches its minimum for any given droplet pair, 
which tends to bring all of the data into alignment 
with lubrication theory toward small $s/R$ in a way consistent with the model in \cite{jeffrey1984}. 

Figure \ref{fig:accelerations} (insets) shows that the $1/s$ scaling given by 
lubrication theory, which holds for small $s/R$, 
holds approximately even at relatively large separations, $s/R \sim O(1)$. 
In this regime of particular interest, 
HI generated \textit{larger} relative accelerations than predicted  
by a factor of about two. 
This factor of two corresponds to a value for $A_c$, 
which is quadratic in the strength of the interaction (Eq.\,\ref{eq:dac}), 
that is about four times larger than predicted and equal to approximately five. 
That is, scaling arguments based on the intrinsic relative velocity predict a threshold relative acceleration 
about three times smaller than observed. 
The threshold corresponding to relative accelerations determined by the initial relative velocity, 
on the other hand, is several hundred times larger than observed. 

	A reduced description in terms of the relative acceleration as presented here 
neglects effects including those of droplet rotations \cite[\textit{e.g.}][]{dhanasekaran2021}, 
the gravitational strength, 
variable droplet size ratios \cite[\textit{e.g.}][]{meibohm2017}, 
nonzero Reynolds numbers \cite[\textit{e.g.}][]{magnusson2022}, 
and non-continuum behaviors in the air at small separations \cite[\textit{e.g.}][]{how2021}, 
as well as the essentially three-dimensional nature of the droplet motions and of the turbulent air 
even at those small scales within which the droplet pair interactions occur 
\cite[\textit{e.g.}][]{zeff2003,bourgoin2018}. 
We expect these effects to change the observed threshold relative acceleration in ways that need to be investigated. 
The striking discriminatory power of the relative acceleration in these cloud-like experiments, 
in spite of its simplicity, calls for examination of its relationship to more complex settings.

\section{Concluding remarks}
\label{sec:conclusions}
The relative accelerations of a droplet pair, 
especially in the range of separations between about one-tenth and one diameter, 
was an indicator of whether droplets coalesced or not in an experiment on droplets settling through quiescent air. 
In this sense, the relative accelerations played a similar parametric role to the impact parameter. 
The relative acceleration may be more useful 
since the impact parameter is difficult to define in practical settings such as clouds 
due to the unsteady three-dimensional flows in the far-field of any given droplet pair. 
In the experiment, which was deterministic, 
the rate at which coalescence occurred was set by the (arbitrary) distribution of initial conditions generated by the experiment itself. 
In more complex settings, these initial conditions are typically drawn from probability distributions 
that are determined by turbulence levels, droplet sizes, and so on, 
such that the outcomes of droplet interactions, we conjecture, 
may be predicted by the probability distribution of relative accelerations exhibited in these settings. 
For instance, the collision efficiency may be defined as the probability of appropriately normalized acceleration differences 
(\textit{e.g.} Eq.\,\ref{eq:dac}) 
exceeding, say, one at surface-to-surface separations of one diameter, or one impact radius. 
To test the utility of this hypothesized indicator of the collision efficiency, 
we suggest that future studies examine the way relative acceleration distributions vary with scale 
and the way they relate to the characteristics of the turbulence and droplets (or particles).

\textit{Acknowledgements:} We are grateful to L.\,Collins, A.\,Dubey, B.\,Mehlig, T.\,Schneider and Z.\,Warhaft for stimulating discussion, 
and to K.\,Chang, D. Hanstorp, and G.\,Magnusson for the beautiful experiment. 




\bibliographystyle{apsrev}
\bibliography{mainarXiv}

\end{document}